\documentclass[a4paper]{article}

\usepackage{itgspeech2025}    
\usepackage{times}            
\usepackage[english]{babel}   
\usepackage[utf8]{inputenc}
\usepackage[T1]{fontenc}      
\usepackage[sort&compress,numbers]{natbib}	
\usepackage{caption}
\usepackage{subcaption}
\usepackage{wrapfig}
\usepackage{makecell}

\usepackage{eqparbox}
\usepackage{float}
\usepackage{lipsum}
\usepackage{multirow}
\usepackage{scalerel}
\usepackage{amsmath,amssymb}
\usepackage{mathtools}
\usepackage{graphicx}
\usepackage{xcolor}
\usepackage[colorlinks=false,pdfborder={0 0 0}]{hyperref}
\usepackage{units}
\usepackage[capitalize]{cleveref}
\graphicspath{{figures/}}

\Crefname{chapter}{Chap.}{Chaps.}
\Crefname{section}{Sec.}{Secs.}
\Crefname{figure}{Fig.}{Figs.}

\newcommand{\argmax}{\operatornamewithlimits{argmax}}
\newcommand\numberthis{\addtocounter{equation}{1}\tag{\theequation}}

\newcommand{\eos}{\ensuremath{\texttt{EOS}}}

\makeatother

\def\ctc{CTC}

\def\fh{FH}
\def\mono{MonoHMM}
\def\trans{mRNN-T}
\def\aed{AED}
\def\aedten{10K AED}
\def\aedoneone{1K AED-1}
\def\aedonetwo{1K AED-2}

\title{A Comparative Analysis on ASR System Combination for \\ Attention, CTC, Factored Hybrid, and Transducer Models}

\author{Noureldin Bayoumi$^{1}$, Robin Schmitt$^{1,2}$, Tina Raissi$^{1}$, 
Albert Zeyer$^{1,2}$, Ralf Schlüter$^{1,2}$, Hermann Ney$^{1,2}$}

\address{
$^1$Machine Learning and Human Language Technology Group, RWTH Aachen University, 
Germany \\
$^2$AppTek.ai, Aachen, Germany \\
Email: \texttt{
		noureldin.bayoumi@rwth-aachen.de \\
		\phantom{33333} \{schmitt,raissi,zeyer,schlueter,ney\}@cs.rwth-aachen.de
	}
}

\begin{document}

\maketitle

\begin{abstract}

Combination approaches for speech recognition (ASR) systems cover structured 
sentence-level or word-based merging techniques as well as combination of model
scores during beam search. In this work, we compare model combination across 
popular ASR architectures. Our method leverages the complementary strengths of 
different models in exploring diverse portions of the search space. We rescore 
a joint hypothesis list of two model candidates. We then identify the best 
hypothesis through log-linear combination of these sequence-level scores. While 
model combination during first-pass recognition may yield improved performance, 
it introduces variability due to differing decoding methods, making direct 
comparison more challenging. Our two-pass method ensures consistent comparisons 
across all system combination results presented in this study. We evaluate 
model pair candidates with varying architectures and label topologies and 
units. Experimental results are provided for the Librispeech 960h task. 
\end{abstract}
\vspace{-0.3cm}
\section{Introduction \& Related Work}

The combination of models in automatic speech recognition (ASR) has long been 
an area of active research, with early approaches like Recognizer Output Voting 
Error Reduction (ROVER). Later, sentence-level log-linear models~\cite{history1,
history2} introduced more theoretically grounded approaches for integrating 
multiple systems, though practical implementations often relied on 
heuristic-based decision-making.

Theoretically, to determine the word sequence with the highest posterior 
probability in an ASR system based on Bayes decision rule ~\cite{Bayes:1763}, 
all possible word sequences must be considered. However, due to intractable 
computational cost, only a subset of hypotheses with scores near the best 
hypothesis is considered, a method known as beam search. The explored search 
space can be represented either as a lattice or via a simple N-best list.\ In 
principle, given a reduced representation of the search space of one model via 
a set of hypotheses, it is always possible to make further decisions by 
involving additional models.\ This concept explores the fact that a hypothesis 
scored purely by one model could be picked up by another one by leveraging 
their complementary strengths.\ There are also confusion network combination 
methods that are motivated by Bayes risk decoding using the Levenshtein 
distance~\cite{hoffmeister2011bayes}.\ 


With the rise of popularity of sequence-to-sequence \\ (seq2seq) models and 
simple 
beam search approaches, different post-hoc hypothesis combination has been 
explored.\ While lattice-based methods were previously explored for system 
combination, the reduced search space in current end-to-end systems and the 
prevalence of fixed beam search sizes have shifted the focus towards N-best 
list-based techniques.\ These approaches go beyond two-pass methods, 
incorporating model loss contributions during training as auxiliary losses and 
combining scores from several models at the frame and label levels during 
decoding.\ One of the key aspect in the recent research is the combination of a 
label synchronous model such as attention encoder 
decoder~(AED)~\cite{bahdanau2016end,chan2016listen} to time synchronous models 
such as 
connectionist temporal classification~(CTC)\cite{graves2006connectionist} and 
recurrent neural network transducers~(RNN-T)~\cite{graves2012sequence}.\ An 
example of this hybrid method is the CTC/AED model, which employs a shared 
encoder and utilizes log-linear interpolation for score fusion during 
decoding~\cite{hori2017joint}.\ This idea is extended by combining CTC, AED, 
and RNN-T during decoding~\cite{sudo2023asr-4d}.\ However, the proposed 
combination has still high time complexity during decoding and the results show 
the redundancy of CTC and RNN-T when combined with AED.\ One prior work 
proposes a hybrid RNN-T/AED model with shared encoder, where the RNN-T model 
produces hypotheses in a streaming fashion which are then rescored in a second 
pass using the  AED model~\cite{sainath2019two}.\ A further direction is 
post-hoc hypothesis combination, it is also possible to integrate a 
deliberation  decoder which attends to both the encoder and the RNN-T 
hypotheses during the second pass~\cite{hu2020deliberation}. Their findings 
showed that rescoring with attention to multiple first-pass hypotheses 
decreases WER. While successful, this approach includes an additional neural 
network for reranking and also does not treat the overlap of different models 
hypotheses explicitly. 
	Furthermore, \cite{sainath2019two,hu2020deliberation} evaluate their
	method only on in-house data which makes it hard to compare to
	other approaches.

In this work, we examine a straightforward model combination method that 
explores the capability of different architectures with different label units 
in producing sequence-level hypothesis that can complement each other.\
We rescore joint list of hypotheses for all model combination results in the 
paper to allow for consistent comparisons between different model combinations.
	While similar to the approach taken in \cite{sainath2019two,hu2020deliberation},
	we use both models for generating hypotheses as well as rescoring while they
	use the RNN-T only for generating hypotheses and the AED only for rescoring.
Using first-pass recognition can give better performance but would make the 
comparison more difficult due to differences in the various decoding strategies.
	For example, the one-pass combination of AED and CTC models can either be
	done time- or label-synchronously, while decoding of the standalone models
	uses either one or the other, depending on the model.
	On the other hand, our two-pass method utilizes the native decoding
	of each individual model to generate hypotheses, followed by a unified
	combination approach which is the same for all combinations. This makes
	our approach more suitable for a systematic comparison.
To our knowledge, there is no prior work that study the model combination for 
different ASR architectures with our proposed method. 
    A critic might question the purpose of combining rather small models in a
	time where huge foundation models demonstrate good performance 
	on their own \cite{radford2023whisper}. 
	We argue that model combination allows us to leverage
	the strengths of different models. For example, under domain
	shift conditions, certain models (e.g. \aed{}), even with more
	parameters and data, face larger performance degradations, while
	the combination with models that are mode robust to domain shift
	can address the problem.
\vspace{-0.3cm}
\section{Model Combination}
\label{sec:modelcomb}
In order to analyze the complementary aspect in the model combination, we 
designed a two-pass model combination method.\ We train different ASR 
architectures and obtain a set of hypotheses that represent the optimal search 
space given the model parameters.\ This means we ensure that our models do not 
have search errors.\ We then combine the set of hypotheses from each pair of 
models and obtain a joint list.\ This joint list is then rescored by both 
models.\ We perform a log-linear combination of the two scores and select the 
hypothesis from this joint list with highest score.\ For a model pair, we form the joint N-best list, rescore with both models, and linearly combine sequence scores with weights $w_1+w_2=1$.\ We pick $w_1$ by grid search on dev-other. We also report an oracle “cheating WER” (best hypothesis in the union per utterance) as an upper bound.\ Our primary objective is to assess how sentence-level score combination can uncover hypotheses with higher scores, that were discarded when considering only the individual model scores.\ We analyze the hypotheses overlap and examine the 
effect of the combination not only on different models but even for two 
trained models of the same architecture.\ We compare our results to the current 
state-of-the-art results in the literature for the LibriSpeech task.\  
\vspace{-0.3cm}
\section{Models}
\label{sec:models}

Bayes decision rule for ASR maximizes the a-posteriori probability of a 
word sequence $W$ of length $N$ given an input acoustic feature sequence $X$ of 
length $T$~\cite{bayes1763lii}.\ We define the inference rule for each of the 
ASR architectures.\ We denote by $h_1^T$ the encoder output.\ We consider a 
generic output label sequence $\phi$ of length $M$ corresponding to $W$.\  This 
consists of either phonemes or byte-pair encoding units~(BPE)~\cite
{sennrich-etal-2016-neural}, depending on the model.\ For the time synchronous 
models we marginalize $\phi$ over allowed alignment sequences following a given 
label topology.\ Let $y_1^T$ denote the blank augmented alignment sequence, and 
denote $s_1^T$ as the hidden Markov state sequence.\ At each time frame we 
access the last emitted output label via the function $a(.)$ as shown later.
\vspace{-0.25cm}
\subsection{Time Synchronous Models}
\vspace{-0.2cm} 
We use four sequence-to-sequence time synchronous models: (1) a Connectionist 
Temporal Classification~(\ctc{})~\cite{graves2006connectionist} and (2) a 
first-order label context transducer model with strictly monotonic label 
topology~(\trans{})~\cite{tripathi2019monotonic,Zhou+Michel+:2022}, (3) a 
first-order label context factored hybrid~(\fh{})~\cite{raissi2020fh,raissi2023competitive}, and (4) 
a monophone hybrid model trained by summing over right and left label 
contexts~\cite{raissi2025right}.\ A general decision rule for \trans{} model is defined in \cref{eq:trans-decode}.
\ We combine an external language model~(LM) with exponent $\lambda$, and 
subtract the internal LM~(ILM) with exponent $\alpha$ using zero 
encoder method~\cite{Zhou+Zheng+:2022}.\ By dropping the dependency $a_{y_{t-1}}
$ and substituting the ILM with a label prior, we obtain the Bayes decision 
rule for \ctc{}.\
\vspace{-0.25cm}
 \begin{equation}\footnotesize
	\underset{W}{\argmax} \hspace{1mm}  \left\lbrace P^{\lambda}_{\text{LM}}(W) \cdot \underset{\scaleto{y_1^T:\phi:W}{8pt}}{\max} \prod_{t=1}^T \frac{P(y_t | a_{y_{t-1}}, h_1^T )}{P^{\alpha}_{\text{ILM}}(y_t | a_{y_{t-1}})}  \right\rbrace \label{eq:trans-decode}
\end{equation} 
The generative diphone \fh{} uses a joint probability of the current phoneme 
state label and its left phoneme context and can be decoded via \cref{eq:fh} by 
subtracting a context-dependent state prior and using an additional alignment 
state transition model with loop and forward with exponent $\beta$.\ Also in 
this case, by dropping the left phoneme context we obtain our \mono{}.\
{\small
\begin{equation} 
	\label{eq:fh}  \numberthis   \footnotesize
	\hspace{-0.5cm}\underset{W}{\argmax} \hspace{-0.5mm}  \left\lbrace \hspace{-0.5mm}P^{\lambda}_{\text{LM}}(W) \underset{\scaleto{s_1^T:\phi:W}{8pt}}{\max} \prod_{t=1}^T \hspace{-1mm} \frac{P(a_{{s_t}- 1}, a_{{s_t}}| h_t)}{P^{\alpha}_{\text{Prior}}(a_{{s_t}- 1}, a_{{s_t}})} P^{\beta}(s_t| s_{t-1})  \right\rbrace 
\end{equation} 
}

\subsection{Label Synchronous Model}
We use standard encoder-decoder architecture with global attention 
mechanism~(\aed{}).\ For a BPE sequence $\phi$ augmented with the 
end-of-sentence~(\eos{}) token, in our model combination experiments we use the 
inference rule defined in \cref{eq:attention-decode}.\ The label $a_M=\eos{}$ 
implicitly models the probability of the sequence length.\ At each step $i$, 
the decoder autoregressively predicts the next label $a_i$ by attending over 
the whole encoder output $h_1^T$.
	We use the length normalization heuristic with exponent $\delta$ to 
	compensate for the well-known length bias problem of AED 
	models \cite{Zhou+Schlueter+:Interspeech2020}.
\vspace{-0.2cm}
\begin{equation} 
	\label{eq:attention-decode} \footnotesize 
	\hspace{-0.3cm} \underset{\{M,a_1^M:W\}}{\argmax} \hspace{1mm}  \left\lbrace  \frac{1}{M^{\delta}} \prod_{i=1}^M  P(a_{i} | a_{1}^{i-1}, h_1^T)  \right\rbrace 
\end{equation}  \vspace{-0.2cm}

%
%

\begin{table}[t]
	\setlength{\tabcolsep}{0.4em}\renewcommand{\arraystretch}{1.1} 
	\caption{Performance of our models trained on LBS 
		960h and evaluated on dev-other and test-other sets.\ We
		report the label unit and context length, as well as the number of 
		epochs for Viterbi~(VIT) and full-sum~(FS) training. All decodings except for 
		AED use a 4gram LM.}
	\vspace{-.3cm}
	\label{tab:baseline-wers}
	{\footnotesize
		\begin{tabular}{|c|c|c|c|c|c|ll|}
			\hline
			\multirow{3}{*}{Model} & \multicolumn{2}{c|}{\multirow{2}{*}{\begin{tabular}[c]{@{}c@{}}AM\\ Label\end{tabular}}} & \multicolumn{3}{c|}{Train}                                                 & \multicolumn{2}{c|}{WER}                                                               \\ \cline{4-8} 
			& \multicolumn{2}{c|}{}                                                                    & \multicolumn{2}{c|}{\#Epochs}                      & \multirow{2}{*}{\#PM} & \multicolumn{1}{c|}{\multirow{1}{*}{dev-}} & \multicolumn{1}{c|}{\multirow{1}{*}{test-}} \\ \cline{2-5}
			& \multicolumn{1}{c|}{Unit}                      & Ctx                                     & \multicolumn{1}{c|}{VIT} & \multicolumn{1}{c|}{FS} &                       & \multicolumn{1}{c|}{other}                     & \multicolumn{1}{c|}{other}                      \\ \hline

			\ctc & \multirow{4}{*}{Phon} & \multirow{2}{*}{0} &  \multirow{2}{*}{0} & 100 & \multirow{2}{*}{74M} & \multicolumn{1}{l|}{6.2} & 6.6 \\ 	\cline{1-1} \cline{5-5} \cline{7-8}
			\mono   &                   &                                       & &  50 &  & \multicolumn{1}{l|}{6.1}&  6.5          \\ \cline{1-1} \cline{3-8}
			
			\fh                    &  &           \multirow{2}{*}{1}              & \multirow{2}{*}{20}     & \multirow{2}{*}{15}  &\multirow{2}{*}{75M}                      & \multicolumn{1}{l|}{5.6}                  & 6.0                                        \\ \cline{1-1} \cline{7-8}
			\trans                 &   &                        &  &  &                       & \multicolumn{1}{l|}{5.8}                  & 6.3                                        \\ \hline
			
			\aed                  & \multicolumn{1}{c|}{10K BPE}                       & $\infty$                                & \multicolumn{1}{c|}{0}    & \multicolumn{1}{c|}{145}   &     98M                  & \multicolumn{1}{l|}{5.3}                  & 5.4                                        \\ \hline
		\end{tabular}
	}
\end{table}

\begin{table}[t]
	\centering
	\setlength{\tabcolsep}{0.9em}\renewcommand{\arraystretch}{.9} 
	\caption{The word error rates~(WER) of two further AED models
		(later referred to as AED-2 \& AED-3) with 1K BPE 
		vocabulary trained on LBS 960h and evaluated on dev-other and test-other 
		sets. Both models are identical except for the random seed used during 
		training.}
	\vspace{-.2cm}
	\label{tab:1k-aeds}
	\begin{tabular}{|c|c|c|}
		\hline
		\multirow{2}{*}{Model} & \multicolumn{2}{c|}{WER}        \\ \cline{2-3} 
		& \multicolumn{1}{c|}{dev-other} & test-other \\ \hline
		\multirow{2}{*}{AED}   & \multicolumn{1}{c|}{5.5} & 5.4  \\ \cline{2-3} 
		& \multicolumn{1}{c|}{5.4} & 5.7  \\ \hline
	\end{tabular}
	\vspace{-.3cm}
\end{table}

\begin{table}[t]
	\setlength{\tabcolsep}{0.9em}\renewcommand{\arraystretch}{1.1} 
	\centering
	\caption{Model combination between time synchronous models using 4gram 
		LM and 10K BPE AED without LM.\ We report the cheating~(cheat) and real 
		WERs.}
	\vspace{-.3cm}
	
	\label{tab:comb-diff-decodings}
	{\footnotesize
		\setlength{\tabcolsep}{0.5em}
		\begin{tabular}{|cc|ccccc|}
			\hline
			\multicolumn{2}{|c|}{Model 1}                                                                                       & \multicolumn{2}{c|}{Model 2}                                                                                                             & \multicolumn{3}{c|}{WER}                                            \\ \hline
			\multicolumn{1}{|c|}{\multirow{2}{*}{Name}} & \multirow{2}{*}{\begin{tabular}[c]{@{}c@{}}Label\\ Unit\end{tabular}} & \multicolumn{1}{c|}{\multirow{2}{*}{Name}}  & \multicolumn{1}{c|}{\multirow{2}{*}{\begin{tabular}[c]{@{}c@{}}Label\\ Unit\end{tabular}}} & \multicolumn{1}{c|}{Cheat.} & \multicolumn{2}{c|}{Real}             \\ \cline{5-7} 
			\multicolumn{1}{|c|}{}                      &                                                                       & \multicolumn{1}{c|}{}                       & \multicolumn{1}{c|}{}                                                                      & \multicolumn{2}{c|}{\makecell{dev-\\other}}                         & \makecell{test-\\other} \\ \hline
			\multicolumn{1}{|c|}{CTC}                   & \multirow{4}{*}{Phon}                                                 & \multicolumn{3}{c|}{\multirow{5}{*}{-}}                                                                                                                                & \multicolumn{1}{c|}{6.2} & 6.6        \\ \cline{1-1} \cline{6-7} 
			\multicolumn{1}{|c|}{MonoHMM}               &                                                                       & \multicolumn{3}{c|}{}                                                                                                                                                  & \multicolumn{1}{c|}{6.1} & 6.5        \\ \cline{1-1} \cline{6-7} 
			\multicolumn{1}{|c|}{FH}                    &                                                                       & \multicolumn{3}{c|}{}                                                                                                                                                  & \multicolumn{1}{c|}{5.6} & 6.0        \\ \cline{1-1} \cline{6-7} 
			\multicolumn{1}{|c|}{mRNN-T}                &                                                                       & \multicolumn{3}{c|}{}                                                                                                                                                  & \multicolumn{1}{c|}{5.8} & 6.3        \\ \cline{1-2} \cline{6-7} 
			\multicolumn{1}{|c|}{AED-1}                 & 10K BPE                                                               & \multicolumn{3}{c|}{}                                                                                                                                                  & \multicolumn{1}{c|}{5.3} & 5.4        \\ \hline \hline
			\multicolumn{1}{|c|}{CTC}                   & \multirow{4}{*}{Phon}                                                 & \multicolumn{1}{c|}{\multirow{4}{*}{AED-1}} & \multicolumn{1}{c|}{\multirow{4}{*}{10K BPE}}                                              & \multicolumn{1}{c|}{4.0}    & \multicolumn{1}{c|}{4.9} & 5.2        \\ \cline{1-1} \cline{5-7} 
			\multicolumn{1}{|c|}{MonoHMM}               &                                                                       & \multicolumn{1}{c|}{}                       & \multicolumn{1}{c|}{}                                                                      & \multicolumn{1}{c|}{4.0}    & \multicolumn{1}{c|}{4.8} & 5.2        \\ \cline{1-1} \cline{5-7} 
			\multicolumn{1}{|c|}{FH}                    &                                                                       & \multicolumn{1}{c|}{}                       & \multicolumn{1}{c|}{}                                                                      & \multicolumn{1}{c|}{3.8}    & \multicolumn{1}{c|}{4.6} & 5.1        \\ \cline{1-1} \cline{5-7} 
			\multicolumn{1}{|c|}{mRNN-T}                &                                                                       & \multicolumn{1}{c|}{}                       & \multicolumn{1}{c|}{}                                                                      & \multicolumn{1}{c|}{3.9}    & \multicolumn{1}{c|}{4.7} & 5.1        \\ \hline
		\end{tabular}
	}
	
\end{table}

\begin{table}[t]
	\setlength{\tabcolsep}{0.1em}\renewcommand{\arraystretch}{1.1} 
	\centering
	\caption{Results for combining similar models on dev and test data. We show the effect of 
		combination for two 1K BPE AED models trained with different seeds.\ We 
		also combine the label posteriors of a monophone hybrid to our \fh{} 
		decision rule using also a 4gram LM.\ }
	\vspace{-.3cm}

	\label{tab:comb-sim-models}
	{\footnotesize
		\setlength{\tabcolsep}{0.4em}
		\begin{tabular}{|cc|ccccc|}
			\hline
			\multicolumn{2}{|c|}{Model 1}                                                                                       & \multicolumn{2}{c|}{Model 2}                                                                                                            & \multicolumn{3}{c|}{WER}                                            \\ \hline
			\multicolumn{1}{|c|}{\multirow{2}{*}{Name}} & \multirow{2}{*}{\begin{tabular}[c]{@{}c@{}}Label\\ Unit\end{tabular}} & \multicolumn{1}{c|}{\multirow{2}{*}{Name}} & \multicolumn{1}{c|}{\multirow{2}{*}{\begin{tabular}[c]{@{}c@{}}Label\\ Unit\end{tabular}}} & \multicolumn{1}{c|}{Cheat.} & \multicolumn{2}{c|}{Real}             \\ \cline{5-7} 
			\multicolumn{1}{|c|}{}                      &                                                                       & \multicolumn{1}{c|}{}                      & \multicolumn{1}{c|}{}                                                                      & \multicolumn{2}{c|}{\makecell{dev-\\other}}                         & \makecell{test-\\other} \\ \hline
			\multicolumn{1}{|c|}{FH}                    & \multirow{2}{*}{Phon}                                                 & \multicolumn{3}{c|}{\multirow{4}{*}{-}}                                                                                                                               & \multicolumn{1}{c|}{5.6} & 6.0        \\ \cline{1-1} \cline{6-7} 
			\multicolumn{1}{|c|}{MonoHMM}               &                                                                       & \multicolumn{3}{c|}{}                                                                                                                                                 & \multicolumn{1}{c|}{6.1} & 6.5        \\ \cline{1-2} \cline{6-7} 
			\multicolumn{1}{|c|}{AED-2}                 & \multirow{2}{*}{1K BPE}                                                  & \multicolumn{3}{c|}{}                                                                                                                                                 & \multicolumn{1}{c|}{5.5} & 5.4        \\ \cline{1-1} \cline{6-7} 
			\multicolumn{1}{|c|}{AED-3}                 &                                                                       & \multicolumn{3}{c|}{}                                                                                                                                                 & \multicolumn{1}{c|}{5.4} & 5.7        \\ \hline \hline
			\multicolumn{1}{|c|}{AED-2}                 & 1K BPE                                                                   & \multicolumn{1}{c|}{AED-3}                 & \multicolumn{1}{c|}{1K BPE}                                                                   & \multicolumn{1}{c|}{4.3}    & \multicolumn{1}{c|}{5.1} & 5.1        \\ \hline
			\multicolumn{1}{|c|}{FH}                    & Phon                                                                  & \multicolumn{1}{c|}{MonoHMM}               & \multicolumn{1}{c|}{Phon}                                                                  & \multicolumn{1}{c|}{4.6}    & \multicolumn{1}{c|}{5.4} & 5.8        \\ \hline
		\end{tabular}
		\vspace{-.3cm}
	}
	
\end{table}
\vspace{-0.2cm}
\section{Experimental Results \& Setting}
\label{sec:exp}
 
 After a short overview of our experimental settings and parameters, we will 
 first report the ASR accuracy of all our models under the Bayes decision rule 
 defined in \cref{sec:models}, following by different model combination 
 experiments.\

\subsection{Setting}
\label{subsec:setting}

We conduct our experiments on the LibriSpeech 960h~(LBS) 
\cite{povey2015librispeech} 
corpus.\ 

For training we utilize the toolkit RETURNN~\cite{doetsch2017returnn}.\ 
Decoding of HMM based models use RASR for the core algorithms, and a recent 
ongoing extension for \ctc{} and \trans{}  decoding~\cite{rybach2011rasr,
zhou2023rasr2}.\ The BPE based models are all decoded in RETURNN as well.\ Our 
experimental workflow is managed by Sisyphus~\cite{peter2018sisyphus}. For more information on training hyper parameters and decoding settings, we refer to an example of our configuration setups
\footnote{\tiny{\url{https://github.com/rwth-i6/returnn-experiments/2025-model-comb}}}.

We use standard sequence level cross-entropy training for \aed{}, \ctc{}, and 
\mono{} models.\ The \mono{} model uses a factored loss with auxiliary 
summation over right and left contexts~\cite{raissi2025right}.\ The remaining 
first-order label context models are trained by first training a zero-order 
label context \ctc{} and posterior HMM models~\cite{raissi2022hmm}.\ We then 
train our models with Viterbi approximation using the fixed path taken by force 
aligning these models for \trans{} and \fh{}, respectively.\ We finish the 
training with a fine-tune phase using the full-sum 
criterion~\cite{Zhou+Michel+:2022}.

For phoneme-based models, we use Gammatone filterbank 
features~\cite{schluter2007Gammatone} with 50 dimensions using 25 
milliseconds~(ms) windows with a 10ms shift. For BPE-based models, we use 
80-dimensional log-Mel filterbank features with the same window and shift.\  
SpecAugment is applied to all models~\cite{park2019specaugment}.\ All acoustic 
models use a 12-layers Conformer encoder with an internal dimension of 
512~\cite{gulati2020conformer}.\ Models with phoneme label unit use a 
downsampling of factor four.\ This factor is increased to six for the BPE-based 
models.\ The alignment models utilize a recurrent encoder with 6 bidirectional 
long short-term memory~(LSTM) \\ \cite{hochreiter1997long} layers having 512 
nodes 
per direction, for a total of $\sim$46M parameters.\ We use one cycle learning 
rate schedule~(OCLR) with a peak LR of around (Viterbi: 8e-4, full-sum from 
scratch: 4e-4) over 90\% of the training epochs, followed by a linear decrease 
to 1e-6~\cite{smith2019super,Zhou+Michel+:2022}.\ The fine-tuning is done on a 
constant lr of 8e-5.\ Our \mono{} model uses auxiliary summation over 
left and right phoneme context~\cite{raissi2025right}, and \aed{} utilizes a 
\ctc{} auxiliary loss \cite{hori2017joint}.\ Following existing setups 
\cite{zeyer2018improved}, we build our systems with either 1K or 10K BPE 
subword label units.\ The \aed{} model uses an LSTM decoder with single-headed 
multilayer perceptron~(MLP) cross-attention \cite{LuongPM15}.\

For decoding, we either use the official LBS 4-gram or a custom Transformer 
\cite{beck2020lvcsr} LM.\ For more information about our 
Transformer LM, we refer the reader to \cite{irie19b_interspeech}. We prefix 
lexical tree based decoding and simple beam search for our phoneme- and 
BPE-based models, respectively.\

\begin{figure}[t]
	\centering
	\begin{subfigure}[b]{\columnwidth}
		\includegraphics[height=0.48\linewidth]{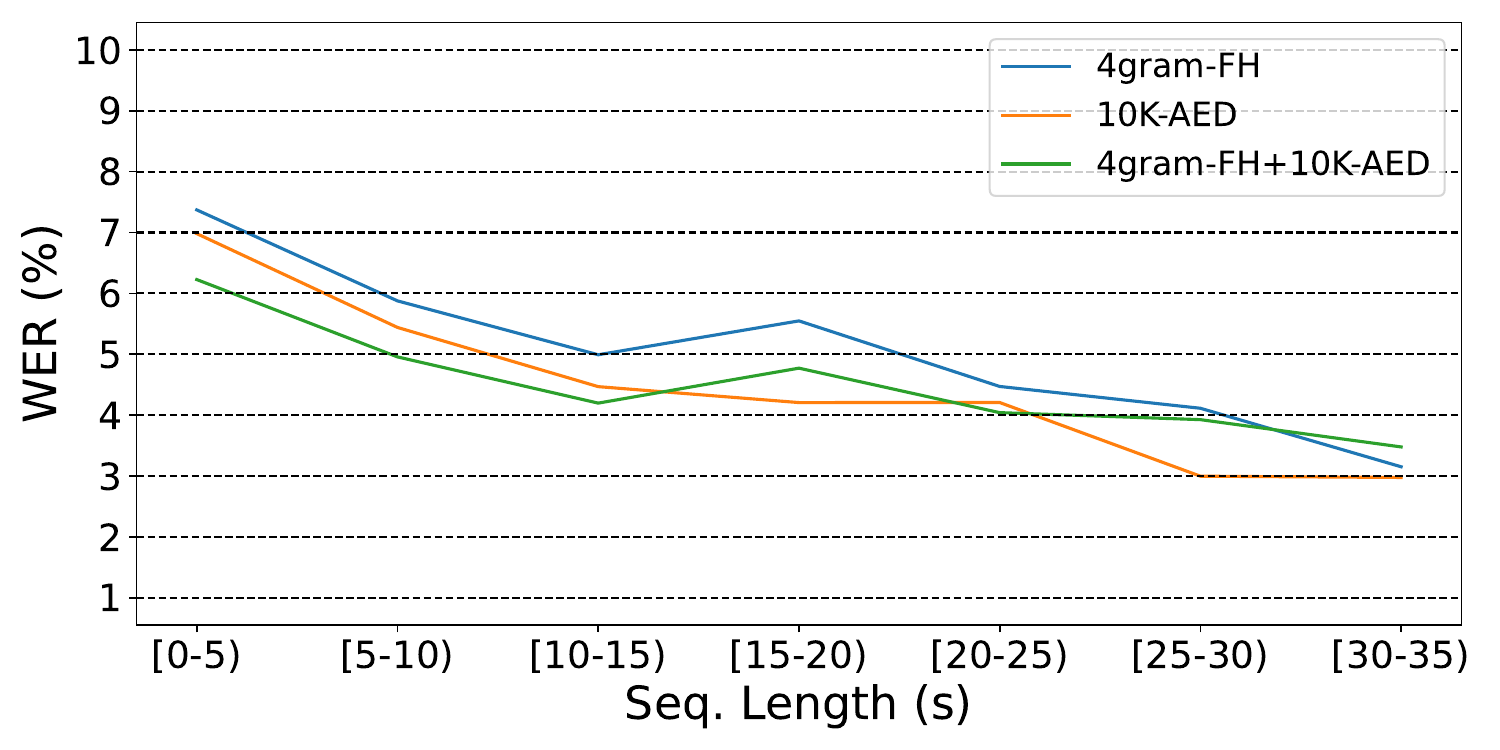}
		\caption{\fh{} and \aedten{}.}
		\label{fig:4gram-fh_vs_10k-aed} 
	\end{subfigure}		
		\begin{subfigure}[b]{\columnwidth}
			\includegraphics[height=0.48\linewidth]{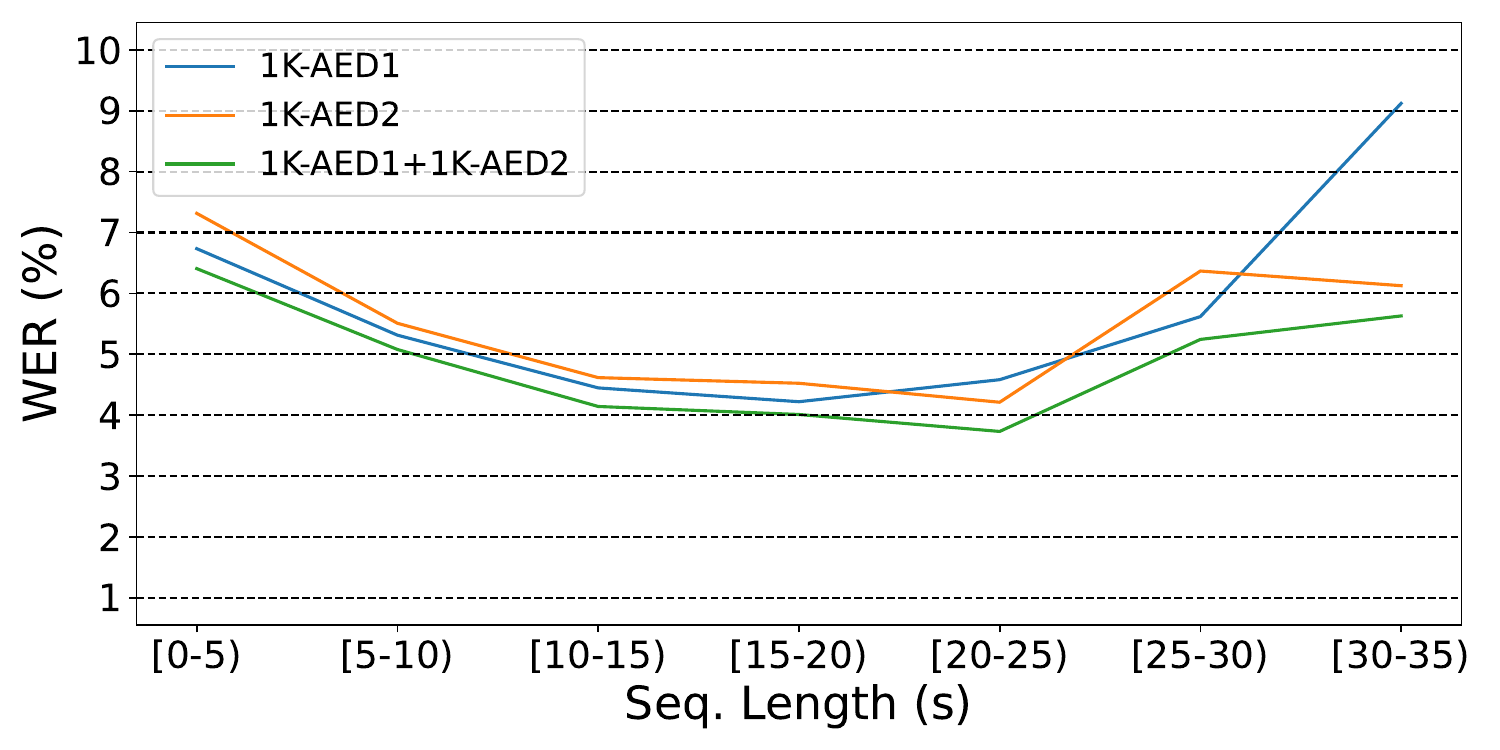}
			\caption{\aedoneone{} and \aedonetwo{}.}
			\label{fig:1k-aed_vs_1k-aed-1234}
		\end{subfigure}
		\vspace{-.6cm}
		\caption[Two numerical solutions]{WERs per sequence length for different models and model combinations.}
		\label{fig:wer-per-seq-len}
	\end{figure}
	\begin{figure}[t]
		\centering
		\begin{subfigure}{.5\columnwidth}
			\centering
			\includegraphics[width = .95 \columnwidth]{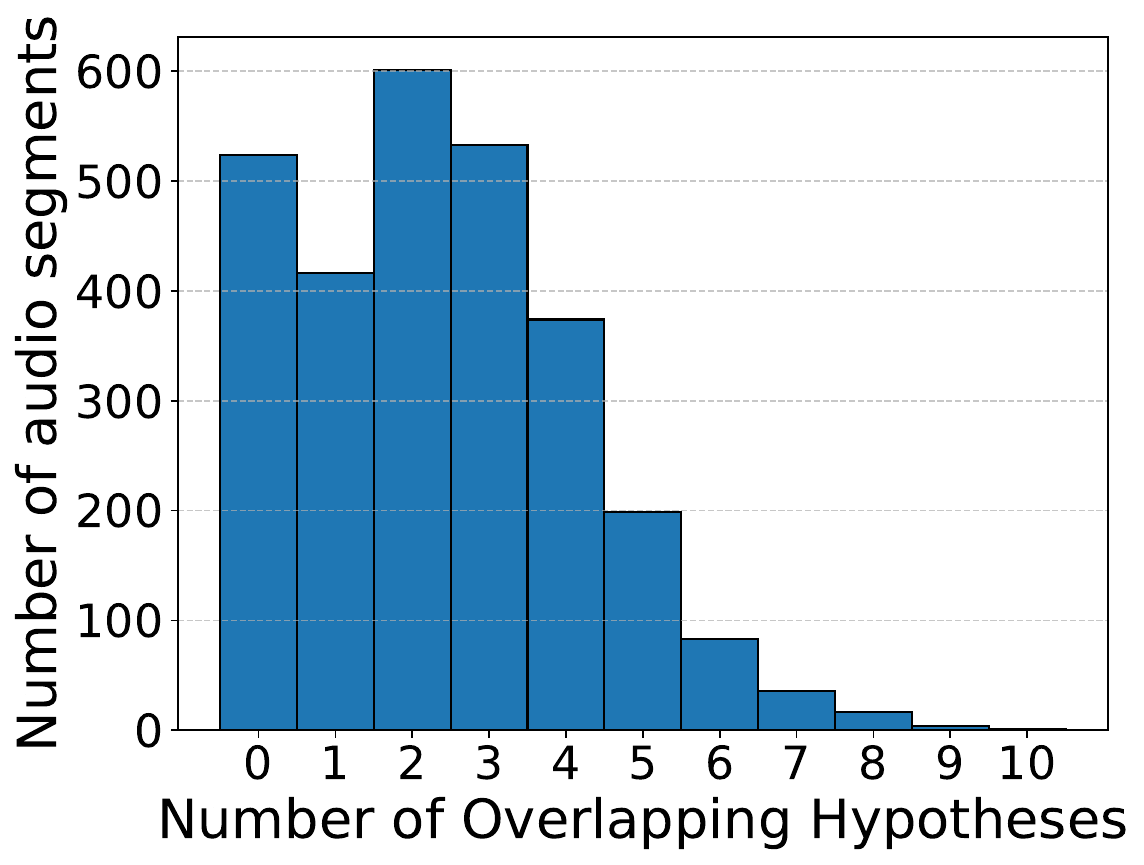}
			\caption{\fh{} and \aedten{}.}
			\label{fig:hyp-overlap-fh-aed}
		\end{subfigure}%
		\begin{subfigure}{.5\columnwidth}
			\centering
			\includegraphics[width = .95 \columnwidth]{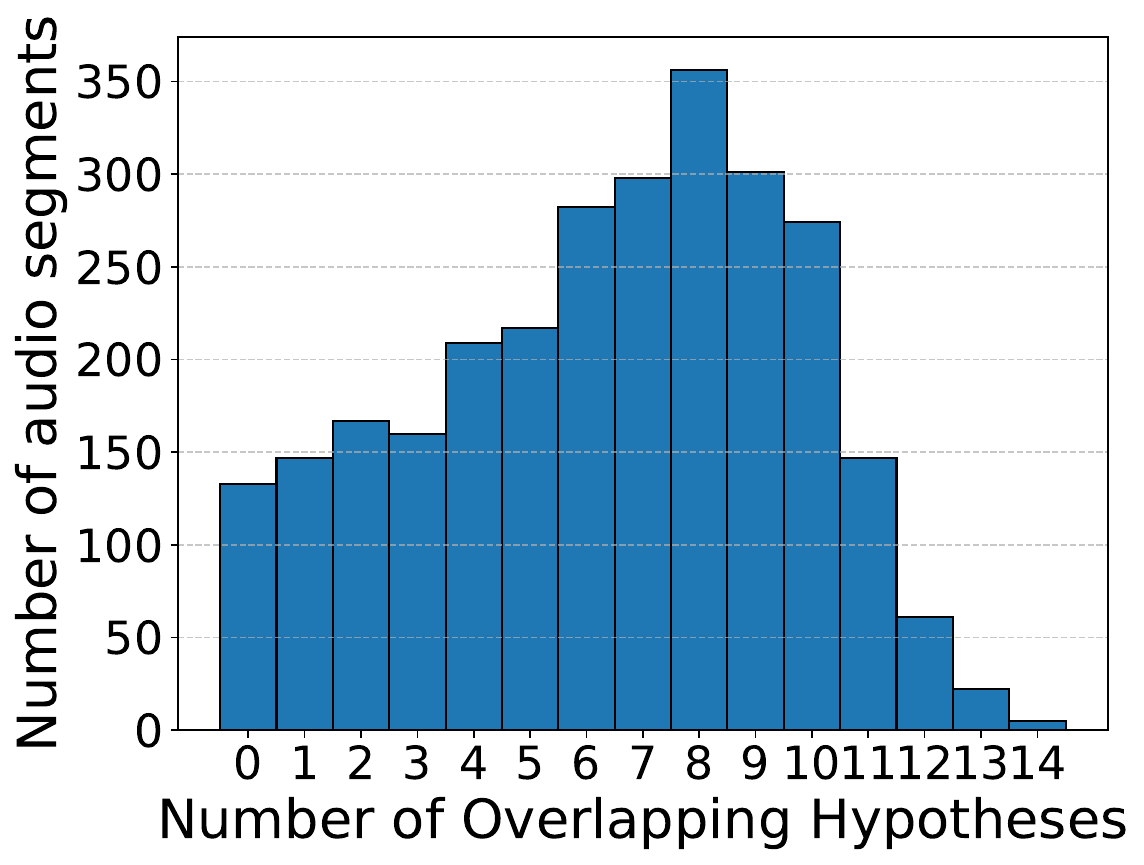}
			\caption{\aedoneone{} and \aedonetwo{}.}
			\label{fig:hyp-overlap-aed-aed}
		\end{subfigure}
		\caption{Hypothesis overlap between two models for N-best lists with N=16.}
		\label{fig:hyp-overlap-n-16}
	\end{figure}

\subsection{Results}
\begin{table}[t]
\setlength{\tabcolsep}{0.1em}\renewcommand{\arraystretch}{1.2} 
\caption{Comparison of the state-of-the-art results on LBS dev-other and 
test-other including model combination approaches. We mention the use of a 
transformer~(Trafo) or 4gram LM where included. In case of \aed{}, when we
use an external LM, we also use internal LM correction.
Results with two digits after the decimal point are not rounded
because the original authors reported them like this.}
\vspace{-.3cm}
\label{tab:final-results}
{\footnotesize
\begin{tabular}{|c|c|c|c|c|c|c|c|c|c|c|c|}
\hline
\multirow{2}{*}{{\scriptsize Work}} & \multicolumn{2}{c|}{AM 1} & \multicolumn{2}{c|}{AM 2} & \multirow{2}{*}{Comb.} & \multicolumn{2}{c|}{LM 1} & \multicolumn{2}{c|}{LM 2} & \multirow{2}{*}{\scriptsize \makecell{dev-\\other}} & \multirow{2}{*}{\scriptsize \makecell{test-\\other}} \\ \cline{2-5} \cline{7-10}
& {\scriptsize Name} & \makecell{Lab. \\ Unit} & {\scriptsize Name} & \makecell{Lab. \\ Unit} & & {\scriptsize Name} & \makecell{Lab. \\ Unit} & {\scriptsize Name} & \makecell{Lab. \\ Unit} & & \\ \hline \hline
\multirow{3}{*}{\cite{sudo2023asr-4d}}
    & AED & \multirow{3}{*}{BPE} & \multicolumn{8}{c|}{\multirow{2}{*}{-}} & 5.6 \\ \cline{2-2} \cline{12-12}
    & \multirow{2}{*}{CTC} & & \multicolumn{8}{c|}{} & 6.9 \\ \cline{4-12} 
    & & & AED & BPE & \multirow{2}{*}{1-pass} & \multicolumn{5}{c|}{-} & 5.2 \\ \cline{1-4} \cline{7-12}
\cite{Kim+Wu+:2023} & AED & BPE & CTC & BPE & & Trafo & BPE & \multicolumn{3}{c|}{-} & 3.95 \\ \hline \hline
\multirow{8}{*}{{\scriptsize Ours}}
    & FH & \multirow{2}{*}{Phon} & \multicolumn{3}{c|}{\multirow{3}{*}{-}} & Trafo & \multirow{2}{*}{Word} & \multicolumn{2}{c|}{\multirow{3}{*}{-}} & 3.9 & 4.2 \\ \cline{2-2} \cline{7-7} \cline{11-12}
    & CTC & & \multicolumn{3}{c|}{} & 4gram & & \multicolumn{2}{c|}{} & 4.9 & 5.2 \\ \cline{2-3} \cline{7-8} \cline{11-12}
    & \multirow{5}{*}{AED} & \multirow{5}{*}{BPE} & \multicolumn{3}{c|}{} & Trafo & BPE & \multicolumn{2}{c|}{} & 3.8 & 4.4 \\ \cline{4-12}
    & & & \multirow{4}{*}{FH} & \multirow{4}{*}{Phon} & \multirow{4}{*}{2-pass} & \multicolumn{2}{c|}{-} & Trafo & Word & 3.5 & 3.9 \\ \cline{7-12}
    & & & & & & \multirow{3}{*}{Trafo} & \multirow{3}{*}{BPE} & \multicolumn{2}{c|}{-} & 3.6 & 4.1 \\ \cline{9-12}
    & & & & & &  &  & 4gram & \multirow{2}{*}{Word} & 3.5 & 4.1 \\ \cline{9-9} \cline{11-12}
    & & & & & & &  & Trafo & & 3.2 & 3.7 \\ \hline

\end{tabular}
\vspace{-.3cm}
}
\end{table}

We present our baseline models in \Cref{tab:baseline-wers}, our results for 
different model combinations in 
\Cref{tab:comb-diff-decodings,tab:comb-sim-models}, and we compare our
best results with the literature in \Cref{tab:final-results}.

Note that we do not report specific combination weights for each model pair. Since we perform log-linear interpolation between sequence-level scores from different architectures — including both label-synchronous (e.g., AED) and time-synchronous (e.g., CTC, FH, RNN-T) models — the absolute values of these scores are not directly comparable across models.\ As such, the resulting combination weights are not interpretable in isolation.\ Nevertheless, we observed that for all combinations involving AED models, the label-synchronous weight was consistently high, typically ranging between $0.960$ and $0.999$.

Classic model ensemble theory suggests that 
the diversity
of combined models is crucial for optimal performance. Our results on dev-other
support this hypothesis since the combination of more diverse models 
(\Cref{tab:comb-diff-decodings}) 
outperforms the combination of similar models (\Cref{tab:comb-sim-models}) 
there. 
For example, the combination of two \aed{} models achieves a 
WER of 5.1\% on dev-other, while the combination of \fh{} and 
\aed{} models achieves a WER of 4.6\%.
However, on test-other,
our results show that we are able to achieve
the same performance (5.1\% WER) by combining two \aed{} models 
(\Cref{tab:comb-sim-models}) 
as by combining a \fh{} and an \aed{} model (\Cref{tab:comb-diff-decodings}). 
This result might be explained by the fact that the WER range between the 
\fh{} and \aed{} models is larger than the WER range between the two \aed{}
models (6.0 \& 5.4 vs. 5.7 \& 5.4).
To further analyze this, we
compare the WER of different models and model combinations for different
bins of sequence lengths in \Cref{fig:wer-per-seq-len}. There, we can see
that, for some sequence lengths, the difference in WER between the two
\aed{} models is as large and sometimes even larger than for the \fh{} and
\aed{} models. Furthermore, in \Cref{fig:hyp-overlap-n-16}, we show the
amount of overlap between the N-best lists of the same two pairs of models
after converting the output labels to words. As expected, the overlap
between the two \aed{} models is much larger than between the \fh{} and
\aed{} models. However, still, we would have expected more overlap in the
former case. For example, for the two \aed{} models, there are still around 150 
audio
segments, where 15 out of the 16 best hypotheses are different. These
two findings show that, while everything, except for the training seed, is
identical for the two \aed{} models, they still produce rather diverse
outputs which helps explain why the combination of them performs 
similarly to the combination of the \fh{} and \aed{} models.

Comparing our baseline models (\Cref{tab:baseline-wers}) with the combination
results in \Cref{tab:comb-diff-decodings} shows that the combination of \ctc{} 
and \aed{} models almost yields the same result (5.2\% WER) as the 
combination of \fh{} and \aed{} (5.1\% WER), even though the standalone \ctc{} 
model 
performs significantly worse (6.6\% WER) than the standalone \fh{} (6.0\% WER) 
model.
A similar observation can be made in case of the \trans{} model.
This suggests that the performance of the combined system is correlated more
with the performance of the better model.
Moreover, in \Cref{tab:baseline-wers}, we can also see that the WER is
dependent on the label context length, with larger context lengths leading to
better performance, which is also true for the corresponding combinations in
\Cref{tab:comb-diff-decodings}.


We also investigated the impact of incorporating a 4-gram LM  for the case of 
\fh{} and 10K BPE \aed{} model combination.\ We first generated the list of 
hypotheses using the transformer LM, and the we rescored this using a 4-gram LM.
\  Interestingly, despite the mismatch between the lattice generation and 
scoring models, the WER on the test-other set of Librispeech 960h improved from 
5.1\% in \cref{tab:comb-diff-decodings} to 4.9\%. This demonstrates that the 
4-gram LM can still provide improvement when applied to transformer 
LM-generated hypotheses list. This suggests that the rescoring  for \fh{} model 
can retain robustness across different LM types, potentially benefiting from 
complementary scoring characteristics. Further analysis could explore how 
different scoring strategies influence final hypothesis selection and whether 
certain model combinations offer advantages in specific scenarios.

In \cref{tab:final-results} we compare our combination method and its effect on 
the word error rate compared to other approaches in the literature.\ We include 
also methods that do not use combination but obtain high ASR accuracy for 
comparison.\ For this purpose, we also further fine-tune our \fh{} model with 
state-level minimum Bayes risk training criterion for one epoch.\
We first compare our two-pass combination of \ctc{} and \aed{} models with the 
one-pass combination of \ctc{}
and \aed{} models from \cite{sudo2023asr-4d} since the corresponding standalone 
models (\Cref{tab:baseline-wers} and \Cref{tab:final-results} top) are
comparable wrt. performance. While our two-pass combination achieves the same 
performance
(5.2\% WER) as their one-pass combination, our standalone models perform better 
which
means that our combination is slightly less effective.
We achieve our best two-pass combination result (3.9\% WER) by combining a \fh 
{} and an \aed{} model,
and using a Transformer, instead of a 4gram LM, for \fh{}. Comparable to this
setting is the one-pass combination of \ctc{} and \aed{} with Transformer LM
in \cite{Kim+Wu+:2023}, which achieves a WER of 3.95\%. 
However, while they
use a Conformer encoder with 17 layers, we use one with only 12 layers
and the same hidden dimension (512).

\vspace{-0.3cm}
\section{Conclusions}

In this work, we investigated the combination of different ASR models using
a two-pass combination strategy which ensures consistent comparisons across
all combination results. We found that the diversity of the combined models
is not necessarily correlated with the performance of the combined system 
by showing that the combination of two identical models can perform as well
as the combination of two very distinct models. Furthermore, we provided
a brief analysis of this finding by comparing the WER of different models
and model combinations for different bins of sequence lengths, and by
analyzing the amount of overlap between the N-best lists of the models.
Finally, we showed that our two-pass model combinations are competitive
with the best single-pass combinations from the literature.
\vspace{-0.3cm}

{\footnotesize 
	\section{Acknowledgments}
This work was partially supported by NeuroSys, which as part of the initiative "Clusters4Future" is funded by the Federal Ministry of Education and Research BMBF (03ZU2106DA and 03ZU2106DD), and by the project RESCALE within the program \textit{AI Lighthouse Projects for the Environment, Climate, Nature and Resources} funded by the Federal Ministry for the Environment, Nature Conservation, Nuclear Safety and Consumer Protection (BMUV), fundingIDs: 67KI32006A. We would like to thank Mohammad Zeineldeen for providing the attention model.}

\newpage 
\small
\bibliographystyle{ieeetr}
\bibliography{mybib}


\end{document}